\documentclass[12pt]{article}
\usepackage{amsmath}
\usepackage{amsfonts}
\usepackage{amsthm, amssymb,multicol}
\usepackage{epsfig}
\font\tenscr=rsfs10 
\font\sevenscr=rsfs7 
\font\fivescr=rsfs5 
\skewchar\tenscr='177 \skewchar\sevenscr='177
\skewchar\fivescr='177
\newfam\scrfam \textfont\scrfam=\tenscr \scriptfont\scrfam=\sevenscr
\scriptscriptfont\scrfam=\fivescr
\def\scr{\fam\scrfam}
\def\centeron#1#2{{\setbox0=\hbox{#1}\setbox1=\hbox{#2}\ifdim
\wd1>\wd0\kern.5\wd1\kern-.5\wd0\fi
\copy0\kern-.5\wd0\kern-.5\wd1\copy1\ifdim\wd0>\wd1
\kern.5\wd0\kern-.5\wd1\fi}}
\def\ltap{\;\centeron{\raise.35ex\hbox{$<$}}{\lower.65ex\hbox{$\sim$}}\;}
\def\gtap{\;\centeron{\raise.35ex\hbox{$>$}}{\lower.65ex\hbox{$\sim$}}\;}

\title{An Alternative Path to the Boundary - the CFT as the Fourier Space of AdS  \vskip .3in}
\author{\large{Ian M. Tolfree\,\thanks{\tt tolfree@pha.jhu.edu}}
\mbox{  }\ \ \\ \\
\emph{\small{Department of Physics and Astronomy}} \\
\emph{\small{Johns Hopkins University}} \\
\emph{\small{3400 North Charles St}}.\\
\emph{\small{Baltimore, MD 21218-2686}}}

\begin{document}
\baselineskip=17pt \pagestyle{plain}
\begin{titlepage}
\vskip -.4in \maketitle

\begin{abstract}

In this paper we shed new light on the AdS/CFT duality by interpreting the CFT as the Fourier space of AdS.  We make use of well known integral geometry techniques to derive the Fourier transformation of a function defined on the AdS hyperboloid.  We show that the Fourier Transformation of a function on the hyperboloid is a function defined on the boundary.  We find that the Green's functions from the literature are actually the Fourier weights (i.e. plane wave solutions) of the transformation and that the boundary values of fields appearing in the correspondence  are the Fourier components of the transformation.  One is thus left to interpret the CFT as the quantized version of a classical theory in AdS and the dual operator as the Fourier coefficients.  Group theoretic considerations are discussed in relation to the transformation and its potential use in constructing QCD like theories.  In addition, we consider possible implications involving understanding the dual of AdS black holes.

\end{abstract}
\thispagestyle{empty} \setcounter{page}{0}
\end{titlepage}

\section{Introduction}

The AdS/CFT correspondence proposed in \cite{Witten:1998qj} is a prescription
for calculating correlation functions for an operator $\scr{O}$ that creates states belonging to a CFT on the boundary.  It is not surprising that a relationship between the two 
theories exits;  they share same symmetry group.   The conjectured equivalence between the two theories, however, is 
surprising.  This paper endeavors to explain the equivalence.  One explanation relies on the fact  that the Fourier
transform of a function defined on anti-de Sitter space is a function defined on the boundary.  To put it another way: the usual quantization procedure of Fourier transforming a function and promoting the coefficients to operators yields operators that create states that only exist on the boundary.   In this regard a CFT naturally arises
as the quantized version of a classical theory in anti-de Sitter
space.   The Fourier transform is
derived following a geometrical technique that was first worked out by Gel'fand and Graev
\cite{Gelfand} (the transform bears their names) and later by N.J.
Vilenkin \cite{Vilenkin}.

This method of deriving a Fourier transformation for functions on the hyperboloid follows from
the analogous derivation of the Euclidean transformation.  One can derive the $n-dimensional$ Euclidean Fourier
transformation in the following way: first calculate the Radon transform of a function
over a plane of dimension $n-1$ and then carry out a $1-dimension$ Fourier
transform over the stack of planes.  The generalization of the Radon transform to a homogeneous space of
constant negative curvature is the Gel'fand-Graev
Transform.  This is the integral transform of a function over a
horosphere, which is the analogy of a plane in AdS.  The $1-dimensional$
Fourier transform reduces this function to its homogeneous
components. 

When quantizing a field in AdS one finds states defined on the
boundary. There is a tower of states at each point, where each rung on the
tower is related to the Fourier transform of the point-horosphere distance function.
This method  confirms the  belief that the bulk function can be viewed as a weighted superposition of boundary states; this is none other than the definition of the Fourier transform.

The outline of the paper is as follows:  first we present a short overview of the two AdS/CFT formalisms to remind the reader what we are trying to point out.  Following this, we briefly review one method of deriving the $n-dimensional$ Euclidean
Fourier transform.  Next we use this method to find the  $n-dimensional$ Fourier
transformation for  Euclidean AdS (Bolyai-Lobachevsky) and take into account group theoretic considerations.  We use the transformation to study a free scalar field in AdS and finally conclude with future considerations.

\section{AdS/CFT Review}

The AdS/CFT relates an $n+1$ dimensional classical theory on AdS to a $n$ dimensional CFT that lives on the boundary.  The two main approaches are that of Witten \cite{Witten:1998qj} and the smearing function technique used in \cite{Hamilton:2006}.  We will start with the former.

The classical bulk action acts as a generating functional for an operator living on the boundary:
\begin{equation}
\label{ac}
    e^{-S} = \langle  e^{\int{d \Sigma \ \Phi0 \ \scr{O}}}  \rangle .
\end{equation}
In practice, one integrates the action by parts:
\begin{eqnarray}
S=\int{dy \partial_M \Phi \partial^M \Phi} = \int{d\Sigma \Phi_0 (\partial_z \Phi )|_{bdy}}  
-\int{dy \Phi \partial^M \partial_M \Phi}
\end{eqnarray}
and then differentiates with respect to $\Phi_0$ in order to generate correlation functions for the operator $\scr{O}$.  $d\Sigma$ is the surface element of the boundary and points perpendicular to the boundary (in the $z$ direction) and hence picks out only the $z$ component of the derivative.  Here the $2nd$ term on the right vanishes since $\Phi$  satisfies the classical wave equation.  The first term on the right would generally be thrown away in flat-space physics since it is a boundary term.  In AdS we cannot simply throw this away because the boundary at infinity can be reached in finite coordinate time so the interior is still accessible from there.  The derivative of $\Phi$ is evaluated at the boundary.  Thus we can write:
\begin{equation}
S=\int{dy \partial_M \Phi \partial^M \Phi} = \int{d\Sigma \Phi_0 (\partial_z \Phi) |_{bdy}}.
\end{equation}
At this point the usual practice is to substitute $\Phi$ in terms of Green's functions that propagate a boundary value of the field to an interior point and then to differentiate with respect to the boundary value in order to generate correlation functions.  If one has extra terms in the Lagrangian a similar technique is applied.

The method involving smearing functions is as follows.  The bulk field is expressed in terms of its behavior near the boundary:
\begin{equation}
\phi(z,x)=\int dx' K(x';z,x) \phi_0(x')
\end{equation}
where $K(x';z,x)$ is known as the smearing function and is calculated from the wave equation.  The fact that the  smearing function is derived from the wave equation hints our interpretation is correct because the Fourier transform is really a plane wave expansion, in hyperbolic space plane waves are just defined differently.   Generically we will refer to irreducible representations of the symmetry group of a space as plane waves.   The correspondence between fields and operators is as follows:
\begin{equation}
\phi(z,x) \longleftrightarrow \int dx' K(x';z,x){ \scr{O}} (x')
\end{equation}
and correlation functions between bulk states are equivalent to superpositions of boundary states weighted by the smearing function:
\begin{eqnarray}
\langle \phi(z_1,x_1) \phi(z_2,x_2) \rangle_{SUGRA} = \int dx_1' dx_2' K(x_1';z_1,x_1) 
\times K(x_2';z_2,x_2) \langle{\scr{O}}(x_1 ') {\scr{O}}(x_2') \rangle
\end{eqnarray}
We now begin with flat space transforms in order to see how smearing functions are consistent with the Fourier interpretation of the CFT.

\section{The Fourier Transform}
\subsection{Flat Space Fourier Transforms}
Before we use integral geometry to derive the AdS Fourier transform we will first use the same techniques to calculate the Fourier transform of a function $f(x)$
defined on $n-dimensional$ Euclidean space. The first step in this method is to calculate the Radon transform of
$f(x)$ over some $n-1 \ dimensional$ hyperplane. The Radon
transform of $f(x)$ over the hyperplane $(x,\xi)=p$ is

\begin{equation}\label{Radon}
\Phi(\xi,p)= \int f(x) \delta((x,\xi)-p) dx
\end{equation}
where $dx$ is the measure of the space and 
$(x,\xi)$  denotes the inner product.  The coordinates of
the function are now $\xi$ and p. Now doing a $1-dimensional$ Fourier
transform over the collection of planes (the p coordinate)
yields
\begin{equation}
f(\alpha \xi)=\int \Phi(\xi,p)e^{ \textit{i} \alpha p} dp
\end{equation}
Setting $\alpha$ equal to 1 and noting how $\Phi$ and p are
defined gives the appropriate n-dimensional Fourier transform
\begin{equation}
f(\xi)=\int \int f(x)e^{ \textit{i}(x,\xi)}dx
\end{equation}
where $dx=dpd\xi$ is nothing more than the volume element of the
space.   This is the n-dimensional Fourier transform of f(x),

\subsection{Fourier Transforms in Anti-de Sitter Space}
\subsubsection{Introduction}
The procedure to calculate the Fourier transform of a function
$f(x)$ in anti-de Sitter space is identical.  The generalization of
the Radon transform is the Gel'fand-Gra\`{e}v transform.  The
generalization of the hyperplane is a horosphere.    The Mellin transform is used in lieu of a Fourier transform; 
one can always
go between the two by a simple change of variables.  Other than
these details the procedure is identical - define something
analogous to a plane, write the integral transformation of $f(x)$
over this plane, carry out a $1-dimensional$ Fourier transform over the set of
planes.   

Before we begin it helps to recall that the whole point of Fourier transforming  is to reduce a function $f(x)$
defined on $n-dimensional$ Euclidean-AdS (Bolyai-Lobachevsky) space into its 
components that transform under an irreducible representation of
the symmetry group of the space. This is what the
usual Fourier transformation does for a Euclidean space function. The
Gel'fand-Gra\`{e}v transformation maps a function $f(x)$ defined
on the hyperboloid to a function $h(\xi)$ defined on the cone
which makes up the points at infinity.  The Mellin transformation reduces this conic function into its homogeneous components \footnote[1]{This usage of the Mellin transformation is the same as in \cite{Gelfand1} and \cite{Vilenkin}}, which can be defined on any contour on the cone.  It is in this way in
which a function in hyperbolic space is reduced into its
homogeneous components.  Once we have carried out this reduction we can promote
the Fourier coefficients to operators and do the usual quantum
mechanics.   

\subsubsection{Geometric Preliminaries}
The distance $r$ between two lines in the embedding space that lie inside the cone $[\xi,\xi]=0$ and pass through the points $x$ and $y$ can be written as
\begin{equation}\label{dist}
\cosh^2(k r) = \frac{[x,y]^2}{[x,x][y,y]}
\end{equation}
where $k$ is the curvature.
When both points lie on the hyperboloid $[x,x]=1$ this reduces to 
\begin{equation}\label{dist}
\cosh^2(k r) = [x,y]^2
\end{equation}
As the point $x$ approaches the cone ($[x,x]=0$ is geometrically a cone and
the points on the cone make up the absolute of the space or the points at infinity), the distance
between the two points diverges.
Points on the cone are zero distance from the origin but infinitely far
away from the hyperboloid that is AdS.  Note that the cone is also
an $n-dimensional$ surface.  $\xi$ will be used to denote a point on
the cone; it is also an $n+1$ dimensional vector.  This fact is important to understanding the significance of the Fourier transform.  In embedding space coordinates the boundary of AdS (the points at infinity) is not simply a plane; it is a surface equal in dimension to that of the hyperboloid.  This is only evident from the embedding space point of view.

From this point of view the embedding space contains hyperboloids and cones which are $n-dimensional$ surfaces in an $n+1 \ dimensional$ embedding space.  The Fourier transform maps a function on the hyperboloid to a homogeneous function on the cone.  Since the components are homogeneous it suffices to know the function on one contour intersecting every generator of the cone once, so any conic section will do; for simplicity pick a circle. One of these circles coincides with the boundary of the hyperboloid in Poincare or plane projective coordinates (AdS coordinate systems are well known, see \cite{Aharony:1999ti} for example).   This is how a CFT can be holographic: scale transformations are equivalent to an extra dimension for a homogeneous function and a homogeneous function on the cone is dual to a function on the AdS hyperboloid.

Returning to horospheres, take a circle centered at a point $a$ and passing
through the point $x$ on the hyperboloid.  Now let $a$ move
off to the absolute of the space.  This leaves a
circle of infinite radius centered on the cone passing through a
point $x$ on the hyperboloid. These great circles are horospheres. A
horosphere is defined by the the point on the boundary at which it
is centered and the point $x$ on the hyperplane through which it
passes.  The equation of a horosphere \footnote[2]{The rescalings $\xi \rightarrow
\alpha \xi$ and $\lambda \rightarrow \alpha \lambda$ produce the
same horosphere so one can always normalize the equation
of a horosphere as desired} is $[x,\xi]=\lambda$.
  So given a point $x$ on the
hyperboloid, each point on the cone defines a horosphere. Consequently the collection of horospheres passing through a point is then equivalent to the set of points on the cone.  

\subsubsection{The Gel'fand -Gra\`{e}v Transform}
We now write down the integral transformation of a function on AdS over a horosphere. This transformation is the Gel'fand -Gra\`{e}v transform and is written as follows:
\begin{equation}
h(\omega)=\int_\omega f(x) d\Sigma.
\end{equation}
$d\Sigma$ is the measure on the horosphere $[x,\xi]=1$ (the
$1$ comes from normalization) and $h(\omega)$ is defined on the
set of horospheres passing through a point.  Each horosphere
though is defined by the point on the cone on which it is
centered.  Since the set of horospheres is equivalent to the points on the cone  we can 
write the transform as
\begin{equation}
h(\xi)=\int f(x) d\xi=\int f(x)\delta([x,\xi]-1)dx
\end{equation}
where $dx$ is the measure of the space and $\xi$ is a point on
the cone.  It is in this way that a function defined on AdS
is mapped to a function on the cone.

We have now written the integral
transform over a hyperplane and now need to do the final $1-dimensional$
Fourier transform.  Note that the integral over the hyperplane is
an integral over the points at infinity.  The goal of this final
transform is to reduce a function defined on the cone into its
homogeneous components.  A function $h(\xi)$ defined on the
boundary is reduced to its homogeneous components by a Mellin
transform:
\begin{equation}
\Phi(\sigma,\xi)= \int_0^\infty h(t
\xi)t^{-\sigma-1}dt
\end{equation}
One can substitute in $h(\xi)$ now and do the t integration. 
This yields
\begin{equation}
\Phi(\sigma,\xi)= \int f(x) [x,\xi]^{\sigma}dx.
\end{equation}
This is the Fourier transform of a function defined in
n-dimensional AdS space.  The $\Phi(\sigma,\xi)$ are the homogeneous
components of $h(\xi)$ of degree $\sigma$.
They are not equivalent to the usual $F(k)$, that is to the usual
Fourier function.  They are the Fourier components, so they are
equivalent to each individual term in the series, i.e. to each
$a_n e^{\textit{i} n x}$.  It is this function which, when promoted to an operator, will create states on the boundary like $\scr{O}$ does. 

The inverse transform for $n=2m+2$ is \cite{Vilenkin}
\begin{eqnarray}
f(x)=\frac{(-1)^m}{(2)(2
\pi)^{2 m+1}\textit{i} \ \Gamma(m+\frac{1}{2})}\int_{a-\textit{i}\infty}^{a+\textit{i}\infty}d\sigma 
\times \int_{[\xi,\xi]=0} \Phi(\sigma,\xi)\delta^{(2m)}([x,\xi]-1)d\xi
\end{eqnarray}
Here the $d\xi$ integral is taken over the cone and $\sigma=
(\textit{i} \rho + a)$.  $a$ can be anything in the range $-2 m+1<a<1$ so long as the integral converges.  We can do these integrals to write the transform in a more lucid fashion by exploiting the homogeneity of $\Phi(\sigma,\xi)$

Since $\Phi(\sigma,\xi)$ is a homogeneous function defined on the cone
it is uniquely determined by its values on any contour which
intersects every generator of the cone once.  One can see this by
using the homogeneity of $\Phi(\sigma,\xi)$ and writing
\begin{equation}
\Phi(\rho,\xi_1,\xi_2,....\xi_n)=\xi_n^{\textit{i}\rho+a}\Phi(\rho,\frac{\xi_1}{\xi_n},
\frac{\xi_2}{\xi_n},......\frac{\xi_{n-1}}{\xi_n},1).
\end{equation}  The right side of the above equation is defined on the
intersection of the cone and the plane $\xi_n$=1, which is just
the sphere $S^{n-1}$. This will be the contour on the cone we
choose to integrate over. We can choose any contour; this one coincides with the boundary of the hyperboloid. This then looks like the boundary of
AdS -- the points at infinity with spherical geometry.

The inverse transform can now be written as (for $n=2m+2$)
\begin{eqnarray}
f(x)=\frac{(-1)^{m}}{2^{2m+1}(
\pi)^{m+\frac{1}{2}}\Gamma(m+\frac{1}{2})\textit{i}} \int_{a-\textit{i}\infty}^{a+\textit{i}\infty}
d\sigma \nonumber \\ \times \frac{\Gamma(\sigma+2m)}{\Gamma(\sigma)}
\int_{\textit{S}^\textit{n-1}}\Phi(\sigma,\xi')[x,\xi']^{-\sigma-2m}d\xi'
\end{eqnarray}
where $\xi'$ denote points on the sphere, which from here on we
will simply call $\xi$.

Since $\Phi(\sigma,\xi)$ is defined on a sphere, one can expand it in
terms of ultra-spherical harmonics.
\begin{equation}
\Phi(\sigma,\xi)=\sum_K a_K(\rho)\Xi_K(\xi)
\end{equation}
where K is a compact notation meaning all of the eigenvalues, e.g.
$K=(l,m_1,m_2...)$.

For $n=6$ (embedding space) the inverse transform becomes
\begin{eqnarray}
f(x)=\frac{(-1)^{2}}{2^{5}(
\pi)^{2+\frac{1}{2}}\Gamma( 2+\frac{1}{2})\textit{i}}\int_{a-\textit{i}\infty}^{a+\textit{i}\infty}
\frac{\Gamma(\sigma+4)}{\Gamma(\sigma)}d\sigma 
\times\int_{\textit{S}^\textit{4}}
\Phi(\sigma,\xi)[x,\xi]^{-\sigma-4}d\xi
\end{eqnarray}

\subsubsection{Quantum Mechanics}
One can now separate the expansion into positive and negative
frequency modes and impose commutation relations on the
coefficients. Let us now analyze what this means. Ignoring the
$\rho$ parameter for a minute, it is evident that the Fourier
transform of a function in AdS is a function defined on the
boundary of the space.  If one were to promote the coefficients to
operators one finds that they create/annihilate states on the
boundary with quantum numbers given by those of the ultra-spherical
harmonics, i.e. states that transform under $SO(5)$.    $\rho$ determines the weight of the representation.
What we see is a tower of states existing on the boundary with
each rung transforming under a different irreducible
representation given by $\rho$.  It is the sum of these states on
the boundary, integrated over the boundary, that is equivalent to
the bulk field.

\section{Group Theoretic Considerations}

\subsection{The Role of the Transform}
We now explain the role of the Fourier transform.
The cone and the hyperboloid are two $n-dimensional$ quadratic surfaces embedded in an $n+1$ dimensional space and consequently  they have the same group of motions.
They have different little groups, however, and are thus different homogeneous spaces.  In other words they have the same group of motions, but different realizations of these motions. 

$AdS_2$ provides a case study.  $AdS_2$ is embedded in $E_{1,2}$ which has the group of motions $SL(2,\textbf{R})$. $AdS_2$ is the homogeneous space $SO(1,2)$ which is the quotient space of $SL(2,\textbf{R})/SO(2)$. 
This realization of the group of motions are the hyperbolic rotations.  $SL(2,\textbf{R})$ has other choices of one  parameter subgroups to serve as the stabilization group.  
One of these is the group Z of lower triangular matrices $\begin{pmatrix} 1 & 0 \\ t & 1  \end{pmatrix} $.  The homogeneous space $SL(2,\textbf{R})/Z$ also has the isometry group $SO(1,2)$.   
This realization of the group of motions, however, is that of the conformal generators defined on the cone.  The Fourier transform serves as a map between these two homogeneous spaces, and thus they are dual. 

\begin{figure}[t]
\hspace{4.315cm}
\epsfig{figure=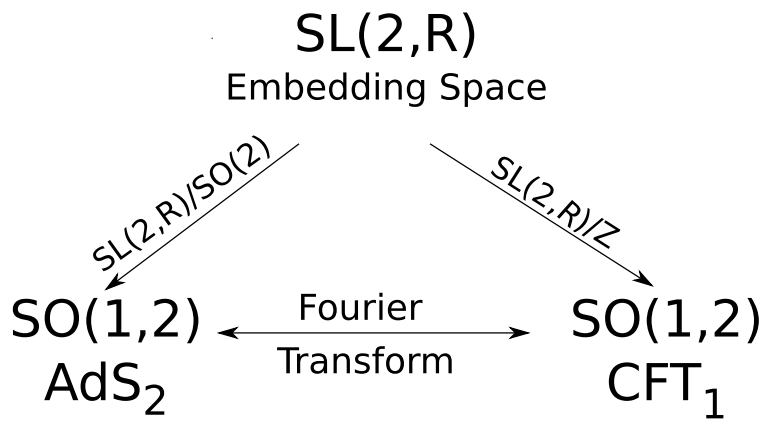,scale=.6}

\caption{\it A schematic showing that two representations of SO(1,2) can be found by taking different quotient spaces of the embedding space.  The Fourier transformation is a means of going between the two representations.  This generalizes for higher dimensional spaces. Because AdS black holes are quotient spaces of AdS they could potentially be included in the diagram with an analogous transform.}
\label{}
\end{figure} 
The kernel of the Fourier transform must then be functions which are irreducible representations of the group.  To find these functions, given the wave operator $\Box$, one needs to expand in eigenfunctions of $\Box$ invariant with respect to the subgroup Z of the cone, the little group.  These will be the analog of the plane waves of the usual Fourier transform.  
One eigenfunction of $\Box$ with an eigenvalue of $-(a^2+\rho^2)$ invariant with respect to the group Z is $z^{a+ i \rho}$, where $z$ is the analog of the $5th$ dimension in Poincare coordinates $(z,\textbf{x})$ and $\rho$ is the weight of the representation.  Under all possible translations, these will transform \cite{Gelfand1} into functions of the form $e^{-(i \rho+a)\tau(x,\xi)}=[x,\xi]^{-(i \rho+a)}$ where $\tau(x,\xi)$ is the point to horosphere distance function.  These are the kernels of our transform derived earlier from geometric considerations.  These are the same solutions to the wave equations that play the role of Green's functions.  It can be clearly seen now how a theory on AdS and a CFT can be dual to one another, and the role the Fourier transform plays in realizing this duality.

\subsection{Recovering the CFT Generators}
Since the Fourier components $\Phi(\rho,\xi)$ are defined on any contour encircling the cone we can pick a spherical contour and expand the Fourier functions in terms of ultra-spherical harmonics.   Because the Fourier components  act on the boundary to create states  that can be represented by spherical harmonics it is easy to see how this process could be overlooked since ultimately it will not change any of the current results of the AdS/CFT.  The integrals and correlation functions will not change.  What is being added is a quantization process.

Let us check the generators now for $AdS_5$.  We started with a function on the hyperboloid invariant under the isometries of $SO(5,1)$ as $15$ rotations.   We then transformed to another function of $\rho$ and $\xi$.  $\rho$ is an invariant, it is equivalent to the energy and  we have seen it is the weight of the representation.  $\Phi(\xi,\sigma)$ is defined on a $4-sphere$ and is consequently $SO(5)$ invariant.  By letting $x \rightarrow \lambda x$ and exploiting the homogeneity of $\Phi(\xi,\sigma)$ we see that the transformation is also invariant under dilatations. To summarize our result, we transformed from the $15$ rotations of AdS to a function that is $SO(5)$ invariant.  The transformation is invariant under dilatations and special conformal transformations (by closure).

These are almost the generators of a CFT.  The only thing left to do is to reduce the $SO(5)$ to $SO(4)$ + translations.    Because any contour on the boundary is equivalent, the contour on the cone coinciding with the boundary of AdS is equivalent to the contour on the cone of infinite radius.  We can pick this and that yields the proper group reduction via a Wigner-Inonu contraction.  The group we want to contract with respect to is the group $SO(4)$.  We will also see what happens to the group representations when we do this.

To carry out the Wigner-Inonu contraction we break up the generators into two  sets:  $J_{1 \nu}=M_{ij}$ and $J_{2 \nu}=M_{0j}$, where $i,j=1,2,3,4,5$ and $J_{1 \nu}$ is the $SO(4)$ subgroup.    Next we redefine the $2nd$ generator  $J_{2 \nu}'= \epsilon J_{2 \nu}$.  The final step is to write down the revised commutation relations between the two sets and take the limit as $\epsilon$ goes to zero.   The result is that an $SO(4)$ subgroup is projected out and we are left with four commuting generators that become the translations, recovering the CFT generators.  We note that $\epsilon$ is not dimensionless; it is equal to $1/R$ where $R$ is the radius of the sphere (contour) we choose on the boundary.    

The last thing to do is to look at how the representations contract.
We can view the ten generators of $SO(5)$ as ten rotations acting on the state $|\textit{l} \  m_1 m_2 m_3 \rangle$  where $l$ is the dimension of the representation.  Contracting the representation involves taking the limit as $l$ goes to infinity and $\epsilon$ goes to zero while keeping the product fixed:

\begin{equation}
\textit{l} \  \epsilon = C
\end{equation}
\begin{equation}
\textit{l} \frac{1}{R}= C
\end{equation}

This is reminiscent of the t'Hooft limit.  When one takes the limit as the $N$ of $SO(N)$ goes to infinity that can be thought of as saying that the dimension of the fundamental representation is going to infinity.  The string coupling constant is related to the radius of the space.  So although we are not taking the limit as $N$ goes to infinity, we are also taking the dimension of the representation to infinity.  It is tempting to say there is a deeper connection, but this will be left to future work.  Since the constant $C$ has units of $\frac{1}{length}$ it seems natural to take this as the QCD  scale \cite{Domoki} and call it $\Lambda$.  

A better way to look at the states is to realize that one can represent $SO(5)$ as $SU(2)\otimes SU(2)$.  So we can write down the states as $|J m_j \rangle \otimes |\Lambda m_\Lambda \rangle$.  The ten generators of $SO(5)$ act on the states as follows: we get two commuting copies of $SU(2)$ that act on each space separately as $ J_i \otimes \textbf{1} $ and $\textbf{1} \otimes \Lambda_i$, where $i=1,2,3$  The remaining four generators are bilinears that act on both states simultaneously: raising both, lowering both, or the two combinations of raising one and lowering the other.  When one contracts and takes $l$ to infinity the two copies of SU(2) decouple (the bilinears are the four $J_{2 \nu}$'s from before that become the translations) and we are left with two copies of SU(2) as our representation of $SO(4)$ and a  knowledge of how the generators act on these states. Now that we know how classical fields correspond to states, and how the generators act on those states, we can start probing deeper into the AdS/CFT.   

Perhaps the most useful aspect of this realization of the AdS/CFT is how it will relate to QCD.  If analytically continued quark states can be mapped to the appropriate representation of $SU(2) \otimes SU(2)$ it should be possible to find the AdS dual of this scenario.  Furthermore, in the contraction scheme the dimensionful parameter $\Lambda$ might help to explain the conformal limit of QCD ($\Lambda=0$) and how when $\Lambda$ is not equal to zero there is a length scale.  It is still unclear if there are any considerations that might compel a specific choice of $\Lambda$.

\section{Quantization in the AdS/CFT}
\subsection{The Generating Functional Method}
If we take this formulation of the AdS/CFT at face value, then when taking the functional derivatives we are literally bringing down a $(\partial_z \Phi(z_b,x))|_{bdy}$.  This is where the correlation functions for the operator $\scr{O}$ come from: this classical function expressed in terms of Green's functions.  These are the classical quantities that perfectly match a quantum theory.  If interaction terms are added, regardless of the strength of the coupling constant, one is still bringing down classical functions expressed in terms of Green's functions. The question left unanswered is where the quantum nature of the AdS/CFT comes from.  
It turns out this formulation of the  AdS/CFT it tells us precisely what to quantize: $( \partial_z \Phi(z_b,x) )|_{bdy}$ for a free field theory. 
 We can view this functional differentiation in terms of correlation functions.  We can also view the functional differentiation in the sense of bringing down a quantum operator that acts on the boundary vacuum.  This quantum operator is just $(\partial_z \Phi(z_b,x))|_{bdy}$ quantized via the usual method:  Fourier transforming and promoting the coefficients to creation/annihilation operators.  
This is reconciled because in \cite{Witten:1998qj} the Green's functions are used to propagate the boundary values.  The derivative does not act on the boundary value of $\Phi$, it acts on the Green's function. 
In fact the weight functions derived here are functionally identical to the Green's functions used in  \cite{Hamilton:2006}.  Their distance $\sigma$ is equivalent to our $[\xi,x]$, which as we saw earlier is nothing but a distance function.  The values of the two distance functions match up for a null, space-like, and time like vectors, confirming our interpretation of the role of the Fourier transform and that the Fourier weights play the role of the Green's function.

  To answer the question of why does it just act on the boundary vacuum is easy:  the quantum states exist only on the boundary, they do not exist in the bulk.  This is what this Fourier transform teaches us.   So we can now correctly view the AdS/CFT in terms of correlation functions or make use of the Fourier transform to go beyond correlation functions and view it in terms of individual quantum states created on the boundary.
\subsection{Smearing Functions}
The  relationship between the Fourier transform and the AdS/CFT becomes apparent when one considers the smearing function formalism of the AdS/CFT.  This formalism states the correspondence as:
\begin{equation}
\phi(z,x) \longleftrightarrow \int dx' K(x';z,x){ \scr{O}} (x').
\end{equation}
From our point of view, we can replace the arrow by an equality.  $\scr{O}$ is the Fourier coefficient $\Phi(\sigma;x)$ that creates states on the boundary and the smearing function $K(x';z,x)$ is related to the Green's function in the usual fashion.  The Green's function is the Fourier weight.  Because the smearing functions are derived from the Green's function it suffices to know the Green's function, which we now calculate from the Fourier transformation for $n=3$ (AdS$_2$):

\begin{eqnarray}
f(x)=\frac{(-1)^{2}}{2^{2}(
\pi) \Gamma(1)\textit{i}}\int_{a-\textit{i}\infty}^{a+\textit{i}\infty}
\frac{\Gamma(\sigma+1)}{\Gamma(\sigma)} cot(\pi \sigma) d\sigma 
\times\int_{\textit{S}^\textit{1}}
\Phi(\sigma,\xi)[x,\xi]^{-\sigma-1}d\xi
\end{eqnarray}
Now let us expand $\Phi(\sigma,\xi)$ in terms of ultra-spherical harmonics as before and change the order of integration, isolating the $\sigma$ integral.
\begin{eqnarray}
f(x)=\frac{(-1)^{2}}{2^{2}(\pi) \Gamma(1)\textit{i}}\sum_k \int_{\textit{S}^\textit{1}} d\xi \lbrace \int_{a-\textit{i}\infty}^{a+\textit{i}\infty}
\frac{\Gamma(\sigma+1)}{\Gamma(\sigma)}d\sigma 
\times a_K(\sigma)[x,\xi]^{-\sigma-1}] \rbrace  \Xi_K(\xi) \
\end{eqnarray}
Notice that the term in $\lbrace \ \ \rbrace$ depends only the variables $\xi$ and $y$ as we are integrating over $\sigma$.  We can do a change of variables from a complex $\sigma$ to a real $\rho$ by means of the following substitution: $\sigma = i \rho + a$.  This integral has a simple pole at $\rho=0$ \footnote[3]{There is still another way to evaluate this integral.  The distance from a point $x$ on the hyperboloid to the horosphere defined by the equation $[\xi,y]=1$ is  $\tau(\xi,y)=log[\xi,x]$.   The transform can now be written in a way more reminiscent of the usual Fourier transform$
f(x)=\frac{(-1)^{2}}{2^{5}(
\pi)^{2+\frac{1}{2}}\Gamma( 2+\frac{1}{2})\textit{i}}\int_{a-\textit{i}\infty}^{a+\textit{i}\infty}
\frac{\Gamma(\sigma+4)}{\Gamma(\sigma)}d\sigma
\int_{\textit{S}^\textit{n-1}}
F(\xi,\sigma)e^{-(\sigma+4)\tau(\xi,y)}d\xi$
.  The ratio of gamma functions can be transformed into polynomials which in turn can be transformed into
 derivatives of tau, as in the usual Fourier transform}.  We are free to choose $a$ so long as the integral
 converges and $a=-\Delta$, where $\Delta$ is the conformal weight, is a good choice.  For odd dimensional embedding spaces this integral has a pole at $\rho=0$ and only $a$ is left.   For even dimensional embedding spaces it seems all terms need to be included.   For $n=3$ embedding space, the sigma integral $\sim [x,\xi]^{\Delta-1} a_K (-\Delta)$.
We can relabel the boundary coordinate $\xi$ and call it $x'$.  Using Poincare coordinates for the bulk, we can also
 rewrite the weight function minus the operator as $G(x'|z,x)$.  The transform then reads:
\begin{equation}
f(x)=\frac{(-1)^{2}}{2^{2}(
\pi) \Gamma(1)\textit{i}}
 \int_{\textit{S}^\textit{1}} dx' \ G(x'|z,x) \ \Phi(-\Delta;x')
\end{equation}
where $\Phi(-\Delta,x')=\sum a_K(-\Delta) \Xi_K(x')$.

The function $G(x'|z,x)=[x,x']^{\Delta-1}$  looks like the Green's function of \cite{Hamilton:2006} where $[x,x']$ is equivalent to their $\sigma$.  Once we know the Green's functions we necessarily know the smearing functions because these are calculated from the Green's functions.  This entails taking the derivative normal to the boundary of the Green's function to find $K(x'|z,x)$, which is precisely what one must do to get the quantum operator $\scr{O}$ in the generating functional formalism.  We have come full circle and can see now how these two formalisms can be resolved by making use for the Fourier transform.   

Because $K(x'|z,x)$ inherits its properties from the Green's function it obeys the desired properties of the smearing functions mentioned in \cite{Hamilton:2006}, such as compact support on the boundary.  We can interpret $\Phi(-\Delta;x)$ as a quantum operator that acts on the boundary vacuum to create states of weight $-\Delta$ that transform under the CFT representation of $SO(5,1)$.  As discussed earlier, the states can be represented by  $SU(2) \otimes SU(2)$.   The states created on the boundary are localized on the boundary but are non-local on the boundary.  This confirms the known AdS/CFT fact that the boundary operators are non-local. 

Without loss of generality one can add a 5th coordinate that coincides with the boundary, $z_b$, to $\Phi(-\Delta;x)$ and rewrite it as $\Phi(-\Delta;x,z_b)$.  This is useful to express  functions defined on a section of the cone and those defined on the hyperboloid in the same coordinates.  $\Phi(-\Delta;x,z_b)$  looks just like the boundary value of a bulk field in addition to an operator that creates conformal states on the boundary.    The difference is in the former case one is interpreting the classical Fourier transform;  if one imposes commutation relations the form of the transform does not change.  Only the interpretation changes which leads us to the latter case.  The fact that $\Phi(-\Delta;x,z_b)$ is interpreted as the boundary value of a field is enhanced by \cite{Hamilton:2006fh} .  There it is pointed how one can pull out the $z$ dependence of the field with degree of homogeneity $\Delta$ and that fields behaving like this near the boundary are the normalizable modes.  One can only do this if the field is homogeneous function in all variables.  In this case $\Delta$ is the degree of homogeneity and the weight of the representation.   This precisely fits our knowledge of how the Fourier functions behave and explains why only fields with this particular fall-off near the boundary are considered: they are not boundary values of a classical field at all.  They are  Fourier components that look like particular field configurations.   In viewing the Fourier coefficients as creation and annihilation operators it is clear why classical fields correspond to boundary operators.  The fact that the degree of homogeneity of the transform is $-\Delta$ is not a problem because of how $z$ is defined.  In Poincare coordinates, $z$ is the inverse of the $5th$ coordinate.  So when the $5th$ coordinate is factored out it is of degree $-\Delta$, meaning that in Poincare coordinates it is of degree $\Delta$, which is what we expect.  This is in line with the normalizable  modes falling off as $z^{\Delta}\phi_0(x)$ near the boundary.  The limits on $a$ come from requiring the $\sigma$ integral to converge, and this range of $a$ corresponds to the normalizable modes of fields allowed in the correspondence.   In addition, any term in the Lagrangian, regardless of coupling, can be written exactly in terms of these $\Phi(-\Delta;x)$ 's as a composite operator.

\section{Conclusions}

An alternative map between a classical theory in anti-de Sitter space and a
CFT on the boundary of the space is found.  The result is that the CFT can be viewed as the
quantized version of a classical theory in AdS.   This becomes evident after considering the Fourier
transform of a function defined on AdS.  The reason a function on AdS is dual to a function defined on the conical boundary is that both spaces are different quotient spaces of the same group, only with different little groups.  The Fourier transform is a map that goes between the two spaces or representations of the symmetry group.   

The conic function can be reduced in dimensionality because homogeneous functions in $n-dimensions$ are equivalent to scale invariant theories in one less dimension.  From the hyperboloid's point of view, in plane-projective coordinates for example, the contour we reduce  the conical function to can always be chosen to be the contour coinciding with the boundary of AdS.  We are thus forced to reinterpret the role boundary values of the fields play as well as required fall-offs in terms of Fourier coefficients defined on the boundary.  This has been made possible by considering both the hyperboloid and cone from the embedding space point of view and allowed us to find the underlying reasons why AdS and a CFT are dual.   

In addition to the QCD considerations already discussed the other main application is to consider what this technique has to say about the AdS/CFT and black holes, this is the subject of current work.    In analyzing the $2+1$ dimensional AdS black hole in \cite{Banados:1992gq} the authors pointed out how AdS black hole solutions could be viewed as quotient spaces of AdS.  In light of our results, it is conceivable that by taking the appropriate quotient space of a CFT one could arrive at the CFT dual of an AdS black hole.  After this, it is hoped that by using the same integral geometry techniques one should be able to arrive at a similar integral transform relating the two theories.  In addition, it should also be possible to arrive at the transformation from geometric considerations only.  A rotating black hole black hole either corresponds to a chemical potential or a rotating Einstein static universe.  In the latter case, one can imagine trying to view this integral geometry set up from this perspective.  In that case, in order to achieve the desired planes to integrate over as part of the Ge'lfand-Graev transform, one must set the boundary points in the integral transform rotating, and herein lies this duality.  In both cases one just needs to define the appropriate planes to integrate over.  This hopefully will teach us about the microstates of AdS black holes.

\section{Acknowledgments}

The author would like to thank J. Bagger, G. Domokos, S.
Kovesi-Domokos, and M. Son for many useful discussions.  This work was supported in part by the US National Science Foundation, grant NSF-PHY-0401513

\end{document}